# Beyond p values: practical methods for analyzing uncertainty in research

12 July 2016

*Michael Wood*
*University of Portsmouth Business School, UK*
*michael.wood@port.ac.uk or MichaelWoodSLG@gmail.com*

## *Abstract*

This article explains, and discusses the merits of, three approaches for analyzing the certainty with which statistical results can be extrapolated beyond the data gathered. Sometimes it may be possible to use more than one of these approaches. (1) If there is an *exact* null hypothesis which is credible and interesting (usually *not* the case), researchers should cite a p value (significance level ), although jargon is best avoided. (2) If the research result is a numerical value, researchers should cite a confidence interval. (3) If there are one or more hypotheses of interest, it may be possible to adapt the methods used for confidence intervals to derive an "estimated probability" for each. Under certain circumstances these could be interpreted as Bayesian posterior probabilities. These estimated probabilities can easily be worked out from the p values and confidence intervals produced by packages such as SPSS. Estimating probabilities for hypotheses means researchers can give a direct answer to the question "How certain can we be that this hypothesis is right?".

**Keywords: Confidence intervals, Estimated probabilities for hypotheses, Null hypothesis significance tests, P value, Statistical inference**

# Contents

# Tables and figures

# Introduction: the problem

Statistical research is usually based on a sample of data with the conclusions being extrapolated to a wider context. The best way to do this is a source of endless debate in the statistical community, but the approach which has cornered the market in many fields of research is the use of null hypothesis significance tests (NHSTs) and p values, despite very strong arguments against this approach going back over 50 years (Morrison and Henkel, 1970: this is an edited collection of readings, some of which were published well before 1970; Nickerson, 2000). There is a review of the problems, and of some of the very extensive literature on these problems, leading to a " Statement on Statistical Significance and P-values" from the American Statistical Association in Wasserstein and Lazar (2016). In medicine, the problems with p values are more widely recognized, and confidence intervals are often cited instead of or as well as p values. These have several advantages (Gardner and Altman, 1986) over p values, but are rarely used in other fields.

My aim in this paper is to clarify the situations in which NHSTs are a reasonable approach, and when and why confidence intervals should be used. I go on to suggest a third approach - giving "estimated probabilities" for hypotheses. The use of the word "probability" in this context is controversial, so I have used the phrase "estimated probability" to acknowledge this, and draw attention to the necessarily tentative status of these probabilities. However, despite the difficulties, it is important to remember that researchers and their audience want to know how certain their results are, and probabilities are the obvious way to express such levels of certainty. So, arguably, statisticians have an obligation to estimate these probabilities by the best available method. If we don't, misinterpretations will fill the void.

It is also, of course, important to remember that there are other sources of uncertainty in research: measures may be inappropriate or inaccurate, samples may be biased, variables may be confounded, data may be fabricated, and so on. But these are not my concern here.

I will start with two very simple examples analyzed using NHSTs. These are fictional scenarios designed to illustrate the issues without submerging the reader in unnecessary details of specific research projects. Example 1 below is designed to be a simplified version of typical research in the social sciences. Example 2 is not typical, but is a convenient vehicle for illustrating the relatively rare scenarios in which NHSTs are a sensible approach, and also provides a very clear illustration of some of their difficulties, and of the core problem facing any conceivable approach to statistical inference.

## Example 1: A Happiness survey

This research involved obtaining a happiness rating (between 0 and 10) from random samples of 10 people in three countries. One of the researcher's hypotheses was that the mean happiness in one of the countries (Happyland) would be higher than in another (Sadland). The results are in Table 1 below, and the data on which the table is based is at
http://woodm.myweb.port.ac.uk/SL/happysurvey.xlsx. (Ten is obviously an unrealistically small sample, but it is helpful to show the contrast with a larger, more realistic, sample below - see Table 3.)

*Table 1. Mean happiness rating in three countries*

| Country | Mean | Sample size |
|---|---|---|
| Happyland | 6.3 | 10 |
| Otherland | 4.0 | 10 |
| Sadland | 6.0 | 10 |

The ANOVA analysis in SPSS (in the Compare means menu option) gives a p value of 0.044, which is just below the conventional cut-off value of 5%. This is a "significant" result, which suggests that there is a real difference between the mean happiness levels in the three countries.

The rationale behind this analysis goes as follows. We imagine a *null hypothesis* that there is no difference in mean happiness ratings between the three countries, and then work out the probability of getting similar results to those obtained, or more extreme results, *on the assumption that this null hypothesis is true*[1]. The resulting probability is the significance level or p value: in this case it is 0.044 or 4.4%. As this probability is low - less than the conventional 5% level - we conclude that the results are unlikely to have occurred if the null hypothesis is true, so the null hypothesis is probably not true.

The researcher's hypothesis is confirmed in the sense that Happyland has a slightly higher rating than Sadland. However, the p value for this comparison (using the Independent samples t test in SPSS) is 0.673[2], indicating a non-significant result. There is a relatively high probability (67.3%) that the results could have been obtained if there really were no difference between Happyland and Sadland. So we cannot reject the null hypothesis of no difference.

## Example 2: A Telepathy experiment

Telepathy is communication which does not rely on any of the known senses: by some unknown method one person becomes aware of what another person is thinking, despite the absence of any opportunity to communicate by means of hearing, vision, touch or any of the other known senses. A few years ago I did an experiment to see if two volunteers - Peter and Annette - could communicate telepathically. I was in one room with a pack of 50 cards, numbered from 1 to 50, which I shuffled thoroughly, and then cut to display one of the cards which I showed to Peter who concentrated hard on it. Then, in another room, I had Annette who at the pre-arranged time, tried to see if she could determine which of the 50 cards Peter was thinking about. She then wrote the number down on a piece of paper. Annette got the card - Number 48 - right. We checked carefully, but there was definitely no way she could have known which card was chosen except by means of telepathy.  (This is, of course, a fictional scenario.)

Cheating - some sort of surreptitious message to Annette - is an obvious explanation here, but if we rule this out , there were only two viable explanations: either Annette was guessing and was lucky, or she was communicating telepathically with Peter. Which explanation is right?

Here the null hypothesis is that Annette was guessing, and the p value, the probability of Annette choosing the correct card by guessing - is obviously 1/50 or 2%.

The p value can be viewed as a measure of the compatibility of the data with the null hypothesis, with high p values indicating the data is compatible, and low p values indicating it is not[3]. In the present case the p value of 2% is under the conventional cutoff of 5%, so this leads to the conclusion that the data is incompatible with the null - guessing - hypothesis, which suggests that the telepathy hypothesis is true.

Most people would think this conclusion is very silly. Although the experimental result is incompatible with the guessing hypothesis in the sense that it is unlikely given this hypothesis, there is still a 2% chance of Annette guessing correctly which still seems more plausible than the outlandish idea that telepathy is at work. The evidence is not strong enough for this to count as serious evidence in favor of telepathy. This illustrates the obvious fact that if we think one hypothesis is almost impossible, the alternative hypothesis (or hypotheses), however unlikely on the basis of the evidence, becomes the sensible choice. This means there can be no rigid rule for statistical inference that fails to take account of these prior beliefs.

To get stronger evidence we need to repeat the experiment (see Table 3 below) or use more cards. Fairly obviously, with a lower p value, the guessing hypothesis becomes less plausible and the telepathy hypothesis becomes more plausible[4].

This argument follows the standard null hypothesis testing procedure closely, except for the rigid cutoff level of 5%. The level of significance obviously needs to be interpreted in the light of the relative plausibility of the hypotheses under investigation. This is in line with the recommendations of most thoughtful commentators.

In order to give the reader a preview of my argument in this article, Tables 2 and 3 show how all three approaches might be applied to our examples. These will be explained in detail in the rest of this article.

*Table 2: Three approaches to analyzing and presenting conclusions*

|  | *Approach 1: Null hypothesis significance tests* | *Approach 2: Confidence intervals (95%)* | *Approach 3: Estimated probability of hypotheses* | *Sample size* |
|---|---|---|---|---|
| *1a\*: Happiness survey with all three countries* | n/a (P=0.044) | Happyland: 5.5 to 7.1 Sadland: 4.7 to 7.4 Otherland: 1.9 to 6.1 | n/a | 10 from each country |
| *1b\*: Happiness survey to compare two countries* | n/a (P=0.673) | CI for difference \*\* (Happyland - Sadland): -1.2 to +1.8 | Estimated prob that Happyland is happier than Sadland =66% | 10 from each country |
| *2: Telepathy experiment* | P=0.02 | n/a | n/a | 1 experiment |

n/a means not applicable according to my argument in this article, although I have included two p values in parentheses because they are the conventional approach which I discuss.
\*   The Happiness survey results come from SPSS[5].
\*\*  "CI for difference" means that we can 95% confident that the mean happiness in Happyland is somewhere between 1.2 units below Sadland to 1.8 units above. Note that this confidence interval refers to the difference between two countries, whereas the confidence intervals in the cell above it refers to confidence intervals for individual countries.

The sample sizes above are all unrealistically small, and it is important to understand how sample size influences results. Table 3 shows the equivalent results for larger samples.

*Table 3: Three approaches to analyzing and presenting conclusions with larger samples*

|  | Null hypothesis significance tests | Confidence intervals (95%) | Estimated probability of hypotheses | Sample size |
|---|---|---|---|---|
| *1a: Happiness survey with all three countries* | n/a (P=0.000) | Happyland: 6.2 to 6.4<br>Sadland: 5.8 to 6.2<br>Otherland: 3.7 to 4.3 | n/a | 400 from each country |
| *1b: Happiness survey to compare two countries* | n/a (P=0.004) | CI for difference (Happyland - Sadland): +0.09 to +0.51 | Estimated prob that Happyland is happier than Sadland =99.8% | 400 from each country |
| *2: Telepathy experiment* | P=1.0 x $10^{-17}$ | n/a | n/a | 10 experiments |

This table mirrors Table 2 except for the sample sizes. For the Happy survey the data comprises 40 copies of the data used for Table 2 so the mean ratings in three countries are identical to those given in Table 1. For the telepathy experiment I have imagined the experiment was repeated 10 times with the same result (Annette choosing the right card).

Before discussing the three approaches, I will briefly review the problem - statistical inference - that they are all designed go tackle, because the jargon of samples and populations makes this appear more straightforward than it actually is.

## Samples, populations and statistical inference

The usual assumption behind NHSTs and confidence intervals is that the data comes from a sample which is *randomly drawn from a wider, often infinite, population*, and the aim of the research is to investigate this wider population. Mathematically this is a very convenient model, but in the real world it may not be obvious what the sample or population are.

In the Happiness survey, the sample is obviously the group of respondents from each country, and the population is the entire population of the country in question. The sample means can be regarded as estimates of the overall population means, and the p values and confidence intervals refer to the certainty with which conclusions can be extrapolated from sample to population. This fits the standard assumption well.

In the Telepathy experiment the nature of the sample and the population are much less obvious. There are 50 cards but these are not a sample drawn from a larger population of cards. The only obvious way of interpreting the word "sample" here is to say there is a sample of one experiment. Similarly there is no obvious candidate for the "population", but we could say that the observed experiment was randomly drawn from an infinite population of hypothetical experiments. If the null (guessing) hypothesis is true 2% of these hypothetical experiments will result in the right answer.

This is a perfectly respectable model, and in terms of making sense of the probabilities it is a helpful one.

In practice, however, this is longwinded and rather forced. It is much easier to talk about the data (the fact that the one experiment resulted in a correct answer) and inferences from the data (about whether telepathy or guessing is the correct hypothesis). This is the meaning of the phrase *statistical inference*: inferences that go beyond simply describing the data. Sometimes the notion of a population is helpful, but sometimes it is not.

Even when there is apparently a clear population, this may misrepresent the true purpose of the research. Let's suppose that each of three countries had a population of just ten. The sample now is identical to the population, so we have not at first sight got a statistical inference problem and p values are meaningless. We don't need them because we have all the information. However, in practice, we may be interested to see if there are any characteristics of the three countries that mean that future inhabitants may also differ in their level of happiness. Here the idea of a hypothetical population of future inhabitants may be useful: the question then is can the present population be regarded as a random sample from this hypothetical population? This is obviously not an easy question, but any conventional statistical inference from the data presupposes the answer "yes". This sort of scenario is actually rather common: we have all the information about the current group of individuals or whatever, but the question of interest is how we can extrapolate our conclusions to a hypothetical group of individuals in a similar situation in the future.

This should alert the reader to the subtlety of many of the concepts in this area. The phrase "statistical inference" is useful as a general term, but it is obviously important to be aware of the nature of the inference. In the Telepathy experiment, for example, we are inferring, from the experimental data (the fact that Annette chose the right card), a conclusion about which hypothesis is the more plausible (telepathy or guessing). In the happiness survey we want to extrapolate the sample results to the entire populations of the three countries, and possibly to future inhabitants as well.

## Approach 1: Testing a null or baseline hypothesis (NHSTs)

The rationale behind the analysis of both examples as NHSTs is outlined above. NHSTs have the advantage that the p value is a (reasonably) well defined probability, and in the Telepathy experiment at least, the way it works is reasonably intuitive. However there are a number of issues to remember when using NHSTs:

### The conclusion from an NHST should never be a firm rejection of the null hypothesis based on a rigid cutoff level

As we have seen in the discussion of the Telepathy experiment, a sensible conclusion depends on the plausibility of the null and alternative hypotheses and should not be made according to a rigid rule.

## The null hypothesis must be *exact* - otherwise you may get a significant result even if the null hypothesis is true

This may seem like a technicality but it actually goes to the heart of why NHSTs can be so misleading to the uninitiated. The null hypothesis in the Happiness survey comparing two countries is that the mean happiness ratings are *exactly* equal, and with a large enough sample even a trivial difference is flagged as significant: e.g. the larger sample in Table 3 (Row 1b) flags the small difference of 0.3 as being significant ($p<0.05$) evidence against the null hypothesis.

A more natural null hypothesis would be that the rating in the two countries are *approximately* equal - say to within one point on the 10 point scale. However, NHSTs can only work with an exact null hypothesis. It might be tempting to assume, informally, this approximate null hypothesis, but then use the exact hypothesis for the computations. Unfortunately, as Table 3 (Row 1b) shows, this would lead to rejecting the null hypothesis even though it is (approximately) true. (Approach 3 does enable us to analyze the approximately equal hypothesis - see Table 4.)

On the other hand the null hypothesis - guessing - in the Telepathy experiment is exact. *Any* discrepancy over a long sequence of trials, however small, from a 2% probability of choosing correctly would be evidence against the guessing hypothesis.

## Normally NHSTs should be supplemented by citing an *effect size*

The significance level tells us nothing about the *size* of the effect: in the Happiness survey the difference between Happyland and Sadland is only 0.3 but this information is *not* indicated by the significance level. This means that it is important to give the size of the effect found in the sample in addition to the significance level . In the Telepathy experiment the obvious measure of the size of the effect is the proportion of times in a sequence of trials that the correct card is chosen.

## The null hypothesis should be credible and interesting

NHSTs focus on a null hypothesis, but if this If this null hypothesis is not credible and interesting getting evidence against it is a waste of time.  The null hypothesis (guessing) in the telepathy experiment is both credible, and interesting in the sense that *any* discrepancy from it would be evidence of an interesting effect.

On the other hand, the null hypothesis of no difference between the means of the three countries (Row 1a in Tables 2 and 3) is of limited credibility and interest. It seems implausible that the three means should be *exactly* the same, and getting evidence against this hypothesis seems of little interest. The p values do not, for example tell us that the outlier is Otherland which is less happy than the other two countries.

## NHSTs are likely to be misunderstood so take care with jargon

NHSTs are widely misinterpreted: the problems have been extensively documented elsewhere (see above). Some of these problems might be reduced with more care over the language used to report NHSTs

The main culprit is probably the word "significant". In ordinary English this normally means "big" or "important" instead of the statistical meaning of signifying a real effect, even if the effect is tiny. This is sometimes exacerbated by phrasing which implies that significance is a property of a hypothesis instead of the strength of the evidence. So we might talk about a "*significant* difference between the mean happiness ratings in Happyland and Sadland" (Table 3, Row 1b), or, worse, we might specify a hypothesis that "the mean happiness rating in Happyland will be *significantly* more than in Sadlland." If the word "significant" is to be used it should refer to the evidence: "significant evidence of a difference ...", and the hypothesis statement should *not* contain the word significant at all.

NHSTs are about a null hypothesis, although this is often not stated explicitly: the p values will be given and the reader left to guess what the null hypothesis is. This is obviously not helpful: the null hypothesis needs to be stated explicitly . The general phrase "null hypothesis" perhaps makes it sound too much of a non-entity while it is in fact the focus of an NHST. The phrase *baseline hypothesis* might be preferable.

For the Telepathy experiment, the result in Table 2 could be cited as "Probability of getting this result by guesswork = 2%" (Table 2, Row 2). For the Happiness survey comparing two countries (Table 2, Row 1b), the result might be "*On the assumption that the overall mean happiness ratings in Happyland and Sadland are equal*, the probability of getting a difference between the mean ratings in Happyland and Sadland as big or bigger than the observed 0.3 is 67.3%." This is longwinded, and could doubtless be improved, but there is surely a case for avoiding neat, concise jargon which can so easily be cited without any appreciation of its meaning.

### There are alternative approaches

Despite the difficulties implicit in the points above, NHSTs are conceptually simple, in the sense that all we have to do is to imagine a null, or baseline, hypothesis and then work out some probabilities on the assumption that this hypothesis is right. When there is an exact, credible and interesting baseline hypothesis, as with the Telepathy experiment, NHSTs do provide a sensible way to analyze the data.

However, there are alternative approaches, some which do produce what NHSTs do not: estimates of the probabilities of hypotheses. I outline what I consider the main two alternatives below. Wasserstein and Lazar (2016) mention a few more alternative approaches, although these are rarely used, and in my view, are too technical for non-statisticians.

In Tables 2 and 3 above, n/a indicates that the approach is not applicable. For example, I think the null hypothesis for the comparison of the three countries in the Happiness survey (Row 1a, Tables 2 and 3) is not sufficiently credible or interesting to justify testing it. Sometimes just one approach will be viable (e.g. Row 1a and 2 in Tables 2 and 3), on other occasions we have a choice (e.g. Row 1b in Tables 2 and 3).

## Approach 2: Confidence intervals

A widely recommended alternative is to present results as confidence intervals (e.g. Gardner and Altman, 1986). In Tables 2 and 3 95% confidence intervals for the mean happiness ratings are given: for Happyland in Table 2 this interval extends from 5.5 to 7.1. These intervals are based on the t distribution which is shown in Figure 1. As the interval is symmetrical and there is 95% confidence between the limits, there must be 2.5% in each "tail". A similar approach could be used to derive 90% (5.6 - 7.0) or 80% (5.8 - 6.8) confidence intervals.

*Figure 1: Confidence distribution for the mean happiness in Happyland (sample of 10)*

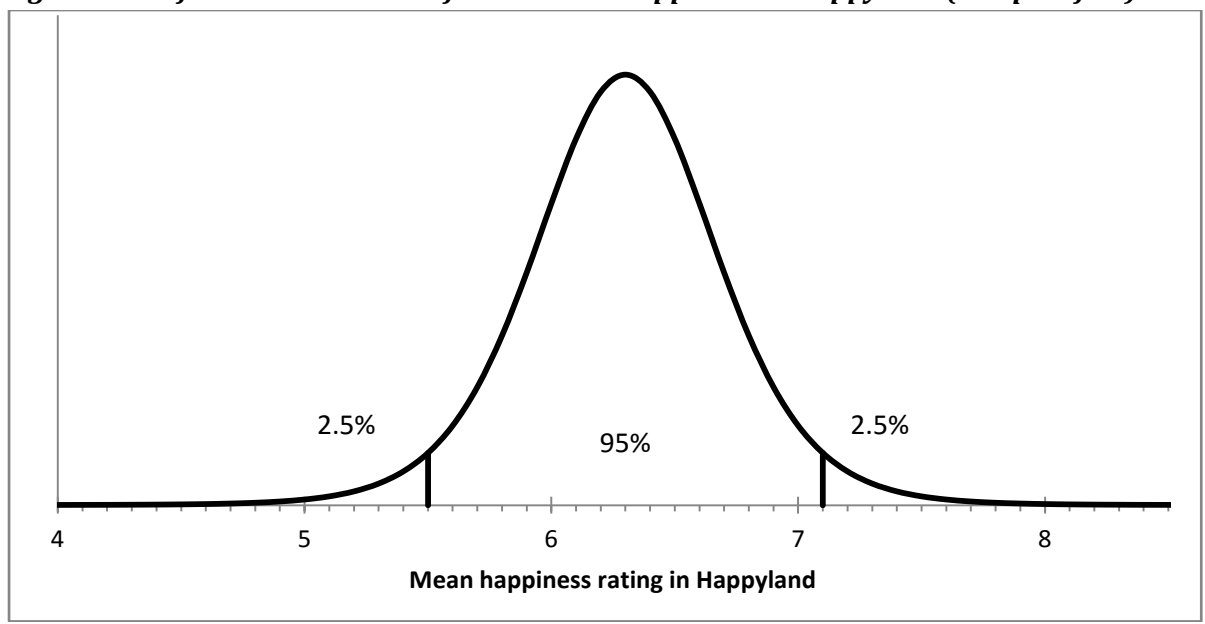

This means that, on the basis of a sample of 10 (assumed random), we can be 95% "confident" that the overall population mean will lie within this interval. On average, this should be true of 95% of such confidence intervals; for the other 5% the overall population mean will lie above or below the interval.

For the larger sample of 400, this interval is narrower, extending from 6.2 to 6.4. (This, of course, refers to the overall population mean, not to individual ratings which, being whole numbers, could not be between 6.2 and 6.4.) This is what we should expect: a larger sample yields a more reliable result which is reflected in a narrower interval.

In practice the word "confident" is taken as meaning the same as "probability", and in rough terms this is reasonable. I will discuss the reason for the distinction below (and argue that it is not a very good reason).

These confidence intervals provide much more information than the p value which simply tells us, in a rather convoluted form, how tenable the hypothesis that all three means are the same is. However, the confidence intervals do not directly address the question of how the means compare with one another. To do this we need a numerical measure of this comparison, and there is no obvious candidate for a measure to compare three means[6].

With a comparison of two means we can simply use the difference of the two means (Row 1b in Tables 2 and 3). For example, in Table 2 with the smaller sample size, the confidence interval for the difference, mean Happyland rating minus mean Sadland rating, extends from -1.2 to +1.8. The bottom of this interval (-1.2) corresponds to a scenario in which the mean happiness in Happyland is 1.2 units *less* than in Sadland, and the top corresponds to the scenario in which Happyland is 1.8 units happier. The sample is small so we can't be sure which country is the happier. With the larger sample, the interval extends from +0.09 to +0.51, indicating we can be reasonably confident that Happyland really is happier.

Confidence intervals for the difference have several advantages over p values. They provide an indication both of the degree of certainty of the result (indicated by the width of the intervals - the width of the interval in Table 2 is 3.0 whereas in Table 3 it is 0.4, indicating the reduction in uncertainty with the larger sample), and of the size of the effect (both intervals are centered on 0.3 which is the difference in mean happiness rating). Furthermore, they avoid the need to set up an exact null hypothesis, and provide a reasonably direct and intuitive way of analyzing results.

Confidence intervals can also be used with regression coefficients and correlation coefficients (Wood, 2005), and, in principle, with any other numerical statistic..

However, confidence intervals are not always possible. There are no obvious natural measures for the size of the effect in Rows 1a and 2 on Table 2,[7] so confidence intervals cannot easily be used here. And they do entail making assumptions which may sometimes not be reasonable, as discussed in the next section.

## Approach 3: "Estimated probabilities" for hypotheses

Approaches 1 and 2 are standard approaches, widely used and supported by software - although Approach 2 deserves to be more widely used and its implementation in SPSS is patchy.

However, in practice, researchers often have a hypothesis whose validity they want to check. Sometimes the evidence might indicate that the hypothesis is definitely true or false, but often this is not the case, so the question arises of how certain we can be - and the obvious way of measuring this is to estimate a probability for the hypothesis. The normal way of achieving this is via Bayes' theorem of probability (see below), but for various reasons (discussed below) conventional statisticians are reluctant to do this, which means that probabilities for hypotheses are rarely presented. However, it is possible to extend the idea of confidence intervals to derive what I will call "estimated probabilities" to acknowledge the difficulties in deriving and justifying these probabilities. Approach 3, then, is a fairly trivial extension to the idea of confidence intervals, which can also be viewed as a crude Bayesian approach as I will explain below. I cannot find any mention of Approach 3 in the literature, which is curious given its obvious usefulness. (It is also worth remembering that there are other approaches to estimating probabilities for hypotheses - such as the direct use of Bayes' theorem.)

There are three, alternative, starting points for Approach 3, all resulting in the similar conclusions. We can start from Bayes' theorem of probability or from p values (see below), but the most obvious starting point is that if it possible to work out intervals corresponding to specified confidence levels, it should be possible to work out confidence levels for specified ranges of the measurement of interest. For example, in Figure 1, the 95% confidence interval for the mean rating of Happyland extends from 5.5 to 7.1. As this interval is symmetrical, this indicates that we can be 2.5% confident that the mean rating is below 5.5, 2.5% confident that the mean rating is above 7.1, and 97.5% (2.5% + 95%) confident that the mean rating is above 5.5. We could use a similar method to work out confidence levels for any range we want (more details below). For example, our confidence level that the mean happiness rating in Happyland is more than 5 is 99.9%, whereas in Otherland it is only 15.2%. These confidence levels refer to our confidence in extrapolating beyond the small sample of 10 to the overall populations of the countries. Although the mean rating in the sample of 10 from Other land is 4.3, the mathematics tells us that there is a 15.2% confidence level that this might have come from an overall population with a mean of more than 5.

We can only use this method when we can work out confidence intervals. For the Telepathy experiment, there is no convincing way of formulating confidence intervals and so this method cannot be used. The obvious way of estimating probabilities for the Telepathy experiment is the direct use of Bayes' theorem - but this is problematic for the reasons discussed below.

We can apply the same method to the *difference* between the means in Happyland and Sadland (Table 2, Row 1b). The mean of the sample rating from Happyland is 6.3 and from Sadland is 6.0 so the difference (Happyland - Sadland) is 0.3. If the difference were zero this would correspond to Happyland and Sadland having equal ratings; if the difference were negative this means that Sadland is happier than Happyland. Table 4 below shows our confidence in various scenarios based on the samples of 10, and on the samples of 400. (The percentages in the table can be worked out either from the mathematics of the t distribution, or from p values or confidence intervals as explained below.)

*Table 4: Confidence levels (estimated probabilities) for hypotheses about how the mean happiness in Happyland compares with Sadland*

| Hypothesis about mean happiness rating | Difference (Happyland - Sadland) | Confidence level or Estimated probability[*] | Size of sample in each country |
|---|---|---|---|
| Happyland ≥ Sadland[**] | ≥ 0 | 66% | 10 |
| Happyland ≤ Sadland[**] | ≤ 0 | 34% | 10 |
| Happyland >> Sadland[***] | ≥ 1 | 17% | 10 |
| Happyland ≈ Sadland[***] | ≤ -1 and ≥ 1 | 79% | 10 |
| Happyland << Sadland[***] | ≤ -1 | 4% | 10 |
| Happyland ≥ Sadland[**] | ≥ 0 | 99% | 400 |
| Happyland ≤ Sadland[**] | ≤ 0 | 1% | 400 |
| Happyland >> Sadland[***] | ≥ 1 | 0% | 400 |
| Happyland ≈ Sadland[***] | ≤ -1 and ≥ 1 | 100% | 400 |
| Happyland << Sadland[***] | ≤ -1 | 0% | 400 |

[*] In the next section I argue that confidence levels can be viewed as estimated probabilities.
[**] The use of the inequalities ≥ and ≤ means that exact equality is included in both categories. In practice this is not a problem because the use of the *continuous* t distribution means that the probability of any exact equality is zero and so can be ignored.
[***] The curly equality sign means *approximately* equal which is defined as being within one unit. The double inequality symbols mean *substantially* more or less in the sense that the difference is more than a unit.

The confidence levels in Table 4 apply to hypotheses about the overall happiness levels in the two countries based on samples of 10 or 400 in each country. All the confidence intervals for the big samples of 400 are close to 0% or 100% indicating a lot of confidence in the hypotheses in question. So, for example with the big sample we have 100% confidence (99.9998% to be more precise) that the happiness levels in the two countries are approximately the same where approximately means within one unit. On the other hand, the small samples provide less confidence as we might expect - only 79%. Obviously with a different definition of "approximately" we would get a different answer: if approximately means "within 0.1 units" the confidence level for the approximately equal hypothesis is 3.6% with samples of 400. If approximately means within 0 units - in other words the hypothesis is exactly equal - its estimated probability is zero.

We can apply this approach whenever the results are analyzed via a numerical measure: this may be a mean, a difference of means, a correlation coefficient or a regression coefficient (see example in the conclusion below).

## Probabilities, confidence levels and Bayes' theorem

Standard statistical theory draws a distinction between "confidence" and probability. However, in practice, confidence statements are usually treated as probability statements, which is what I advocate here. All probabilities are estimates but some, like those derived from confidence intervals, are rougher estimates than others, so I will refer to them as "estimated probabilities". Fairly obviously, any attempt to extrapolate empirical results beyond the data on which they are based, must be hedged in by assumptions and approximations, so these estimated probabilities should always be interpreted with a certain degree of suspicion. The rest of this section amplifies and explains this stance.

According to the dominant ("frequentist") version of statistical theory, the confidence levels in confidence intervals (95%, 90% or whatever) are *not* probabilities. The reason, using the confidence interval for the mean rating in Happyland based on a sample of 10 as an example, is that there just one true mean (although we don't know what it is) so probabilities are irrelevant. Probabilities apply to uncertain *events*, like whether a coin lands on heads or tails, *not* to hypotheses which are either true or false. On the other hand confidence intervals are obviously useful, so a story is needed to make sense of them without admitting that probabilities can be applied to hypotheses. The usual story behind a 95% confidence interval goes something like this: if the procedure that led to the confidence interval were repeated a large number of times, then in about 95% of these repetitions the interval would include the true population mean and in the remaining 5% it would not. The probability of 95% here refers to the *event* of the true mean being inside an interval, not a hypothesis about the true mean.

This story seems to me both dubious (how do we know what would happen if we repeated the procedure a large number of times?) and pointless because, unlike many statisticians, most people have the imagination to make sense of the probability of a hypothesis. Furthermore, if the hypothesis is true in 95% of repetitions of an experiment or survey, this seems a sensible way of estimating a probability for the hypothesis. This is perhaps implicit in Nickerson's (2000: 279) description of the distinction between confidence and probability as "subtle".

Accordingly I will call these confidence levels "estimated probabilities". In practice all probabilities of real world events are estimates: the probability of 50% for a coin landing heads is true of an idealized coin, but a real coin may be slightly asymmetric or it may occasionally land on its side, so as a description of a real coin 50% can only be an estimate. However it is likely to be a fairly accurate estimate based on some reasonably plausible assumptions. Probabilities applied to hypotheses, however, are likely to be far less reliable as estimates: hence my use of the word "estimated".

It is important to view these estimated probabilities with a degree of skepticism. The underlying problem is best illustrated by the Telepathy experiment. As we saw above, if we think that telepathy is a very unlikely hypothesis we would require stronger evidence than if we considered it more plausible. If, for example, we think telepathy is impossible, then we would assume that Annette got the right card by guessing, and our estimated probability for the telepathy hypothesis would still be zero. *This means that any rule for assigning a probability to the telepathy hypothesis needs to take these prior beliefs into account.*

This can be done by means of a theorem of probability called Bayes' theorem. This allows us to take account of prior beliefs, as "prior" probabilities, and work out "posterior" (after taking account of the evidence) probabilities. This is a basis of the Bayesian approach to statistics. The difficulty, and the reason Bayesian statistics is not widely used, is that sensible estimates of prior probabilities may be difficult to obtain. Often these estimates are largely subjective, which means that the final conclusion depend on these subjective judgments - which is viewed a problem by mainstream statisticians.

However, sometimes it is possible to make some fairly neutral assumptions about prior probabilities. With the confidence interval for Happyland, if we assume that all values on the horizontal axis of Figure 1 are *equally likely*, then the Bayesian equivalent of a confidence interval (a credible interval) is *identical* to the confidence interval (Bolstad, 2007, p. 244). From the Bayesian point of view, an important implicit assumption behind confidence intervals, and estimated probabilities derived from them, is that all values on the axis are equally likely. Similar conclusions apply to results based on other statistics like correlation and regression coefficients (Bolstad, 2007, p. 280). There is a general argument explaining how the assumption of equal prior probabilities can be used to derive a posterior probability distribution in the Appendix of Wood (2014).

It is instructive to apply a similar method to the Telepathy experiment. We could assume that the telepathy and the guessing hypothesis are equally likely: Bayes theorem, and, with luck, the reader's intuitions, then tell us that the telepathy hypothesis is 50 times as likely as the guessing hypothesis, so the estimated probability of the telepathy hypothesis is 50/51 or 98%. Most people would consider this a very silly conclusion: in Bayesian terms their prior probability for the telepathy hypothesis is far less than 50%. (If the prior probability were 1%, for example, the posterior probability for the telepathy hypothesis works out as 34%[8].) This should make us cautious about assuming equal prior probabilities. Making unwarranted, implicit assumptions about prior probabilities is a major argument behind Ioannidis's (2005) disturbing claim that, in the medical field, "most published research findings are false".

## Methods of computing estimated probabilities for hypotheses

As we have seen, estimated probabilities for hypotheses may be derived from the mathematics behind confidence intervals, and from Bayes' theorem. However, in practice, researchers use computer packages like SPSS to work out their results, and it is possible to derive some estimated probabilities for hypotheses from the p values and confidence intervals produced by such packages. I'll start with p values because this is the most straightforward route.

### From *p* values

The p value for the comparison between Happyland and Sadland on the basis of samples of 10 is 0.673 which I will write as 67.3%. Conventionally p values are written as fractions, whereas confidence levels are written as percentages: I will work in percentages for estimated probabilities, because this seems more intuitive.

The p value of 67.3% is related to the 100% - 67.3% = 32.7% confidence interval by the fact that one end of the confidence interval is where the difference of the two means is zero (Figure 2). The reason for this follows from the way confidence intervals are constructed[9].

*Fig. 2. Confidence / Estimated probability distribution for the difference between mean Happyland rating and mean Sadland rating*

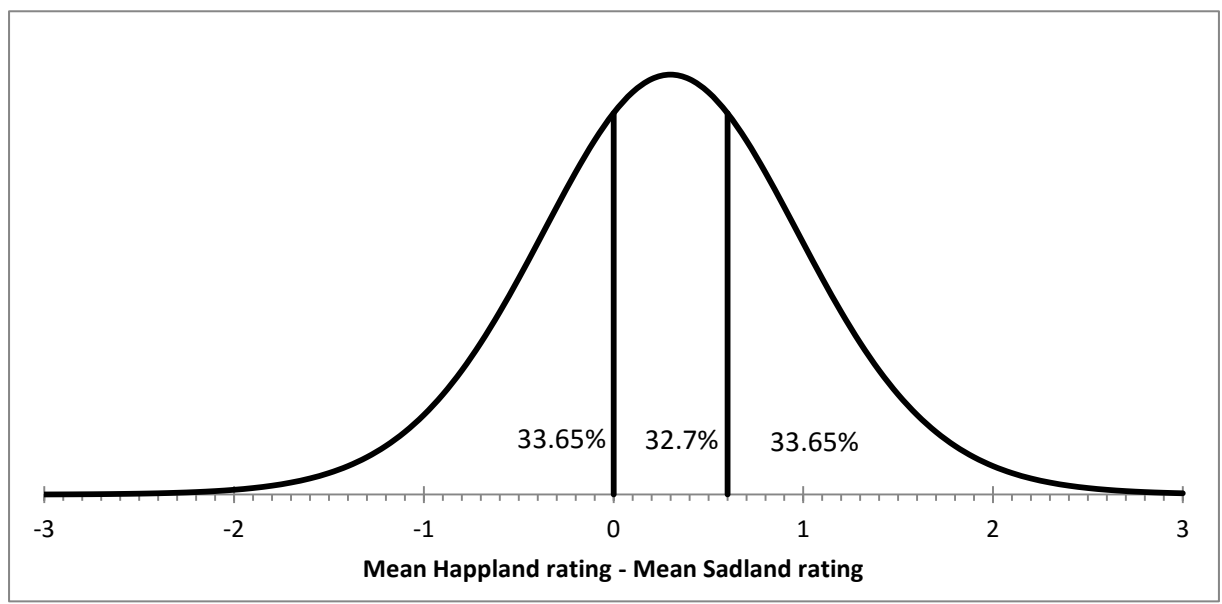

It is obvious from Figure 2 that the Estimated probability of the difference being more than 0 - corresponding to Happyland being happier than Sadland - is simply 32.7% + 33.65% = 66.35%. This can also be worked out as 100% - 67.3%/2. For the larger samples in Table 3 the corresponding estimated probability is 99.8%. The general formulae are in Table 5 below.

*Table 5. Formulae for converting p values to estimated probabilities for hypotheses*

> *Suppose we have a (two tailed) p value for a statistic (e.g. a difference between two means, or a correlation or regression coefficient), and*
>
> If the value of the statistic from the sample is positive (as in this example where the observed value is +0.3), then
>     Estimated probability that population value of statistic is positive = 1 - p/2
>     Estimated probability that population value of statistic is negative = p/2
>
> If the sample value is negative, then
>     Estimated probability that population value of statistic is positive = p/2
>     Estimated probability that population value of statistic is negative =1 - p/2
>
> These formulae can easily be adapted if the information we are given about p is an inequality. For example if p < 0.1% then the first equation above becomes
>     Estimated probability that population value of statistic is positive > 99.95%

**From confidence intervals**

The estimated probability for the mean happiness rating in Otherland being more than 5 is 15.2%. This can be worked out using mathematics of the t distribution. Alternatively it can be worked out by trial and error from the confidence intervals produced by SPSS, although this does require a bit of patience. We need to find a confidence level which makes the top end of the interval about 5. I tried 95% (giving a confidence interval of 1.9 - 6.1), 80%, 65% and so on, until eventually I homed in on 69.5% (3.0 - 5.0). This means that the top tail of the distribution (similar to Figure 1) must be (100%-69.5%)/2 or 15.2%, which is the required estimated probability.

# Conclusions

I have looked at three approaches to analyzing uncertainty in research: these are illustrated by the results summarized in Tables 2, 3 and 4. There are some recommendations about when the three approaches should be used in the abstract.

The most widely used approach is the testing of null hypotheses leading to p values (or significance levels). Sometimes (e.g. the Telepathy experiment described above) this may be sensible, but often it is a deeply flawed approach. To take an example from the recent research literature, Mitsuhashi and Min (2016, P. 295)[10] write:

Firms are more likely to form alliances with others with which they have prior alliances ($p < 0.001$) and shorter distance ($p < 0.001$), co-participate in the same consortium ($p < 0.001$) and share a greater number of common third parties ($p < 0.001$). Membership in different consortiums, on the other hand, reduces the likelihood of alliance formation ($p < 0.001$). Hence, the present findings are consistent with those in previous research.

I would replace "p<0.001" by "Estimated probability > 99.95%" (using the formulae in Table 5 above, remembering that the use of phrases like "have prior alliances" implies that the sample value for the measure of this property was positive).

The purpose of the paragraph cited above is to tell the reader how likely these hypotheses are based on the evidence collected; stating the results in terms of estimated probabilities gives the conclusions directly rather than circuitously by p values referring to an unstated, hypothetical null hypothesis.

Sometimes there is no satisfactory way of estimating probabilities (e.g. the Telepathy experiment above), but when it is possible to work out confidence intervals, estimated probabilities for hypotheses can be derived from these, or from p values. We could use the phrase "confidence levels" instead of estimated probabilities, but this seems an unnecessary complication because they are, reasonably, interpreted as probabilities. However, the word "estimated" is important to remind us that these probabilities must always be regarded as tentative because they are based on assumptions which may sometimes not be entirely sensible.

The other approach is to cite confidence intervals for the parameter of interest (see Tables 2 and 3), as is often done in medical articles. The parameter might be a mean, the difference of two means, or a regression or correlation coefficient. This has the advantage of indicating the size of the effect and the likely margin of error: the confidence interval for the difference of the two means in Table 3 (0.09 - 0.51) makes it clear that, although we can be confident the difference is positive (i.e. Happyland is happier than Sadland), it also makes it clear that the difference is small (bearing in mind the scale of measurement goes from 0-10). Mitsuhashi and Min (2016) use logistic regression which does not yield an intuitive measure, so this approach is not really appropriate here. The word "confidence" is well established and unproblematic in the context of an interval, but we could use the phrase "estimated probability" here too, which has the advantage of reminding us of the tentative nature of these intervals.

There is, of course, no reason why two or more methods of analysis should not be presented. If convention dictates that p values must appear, these could be presented as well as estimated probabilities.

## *Endnotes*

[1] There are different ways of defining the precise null hypothesis and different ways of working out this probability (ANOVA being one of them), so slightly different answers are possible. The Kruskal-Wallis test (in the SPSS Non-parametric tests option) gives p = 0.121, and "Retain the null hypothesis" as the "decision".

[2] This is a two tailed p value because we count results in either direction as being extreme.

[3] According to the ASA Statement on Statistical Significance and P-values (Wasserstein and Lazar, 2016) "P-values can indicate how incompatible the data are with a specified statistical model" which leads to the convoluted statement that "the *smaller* the p-value, the *greater* the statistical *incompatibility* of the data with the null hypothesis ..." (my italics). It is surely better to explain one

quantity (p value) by another quantity that rises as p values rise (compatibility) rather than one which changes in the opposite direction (incompatibility).

[4] When teaching about significance tests I usually use a version of this example as a lecture demonstration. I ensure a positive result by enlisting a collaborator who knows the card that will be chosen, but even with my low level of acting skills, some people in the audience often do think the experiment is genuine. I then ask them to imagine that the experiment has been repeated with the same result: an informal count shows that the number supporting the Telepathy hypothesis increases as the p value decreases.

[5] The ANOVA results and the table of means in Table 1 come from Compare means - Means. However, this option does not offer confidence intervals: for the individual countries these come from Descriptive statistics - Explore, and for the two country comparison from Compare means - Independent samples t test.

[6] Any NHST (e.g. ANOVA) will use such a measure, but the chosen measures are not particularly intuitive ones.

[7] For the telepathy experiment it would be possible to treat the long run probability of Annette guessing correctly as a measure. Then, a standard confidence interval argument might suggest that the observed data gives a confidence interval extending from 5% to 100%. But this is a rather dubious argument and has none of the intuitive transparency of the NHST argument in this case.

[8] Posterior probability = 1%*100% / (1%*100% + (100%-1%)*(1/50)) .This is Bayes' theorem in Excel format. To see what happens with another prior probability, simple change 1% for the new prior probability.

[9] It is sometimes assumed that if a result is *not* significant at the 5% level of significance, this is equivalent to 95% confidence that the null hypothesis is true. In this case the significance level is 67.3%, so the equivalent confidence level would be 100%-67.3% or 32.7%. This is wrong because the null hypothesis is an exact hypothesis so the probability of its being true is zero or very small. However, we can say, in this example, that we can be 32.7% confident that the true population difference of the means will be somewhere between 0 and 0.6.

[10] I chose this article because it had a convenient paragraph citing p values for hypotheses which make sense without reading the rest of the article. However, the table on which they are based suggests that the p values cited are wrong: they should all be 0.01 which means that the estimated probabilities are all > 99.5%.